\documentclass[twocolumn,prb,superscriptaddress,showpacs]{revtex4}

\usepackage{amsmath,dcolumn,graphicx}

\newcommand{\ucupdx}{UCu$_{5-x}${Pd}$_{x}$}

\newcommand{\ucu}{UCu$_{5}$}
\newcommand{\upd}{UPd$_{5}$}
\newcommand{\xlow}{UCu$_{4}$Pd}

\begin{document}

\title{Effects of lattice disorder in the UCu$_{5-x}${Pd}$_{x}$ system}

\author{E. D. Bauer} \affiliation{Department of Physics and Institute
For Pure and Applied Physical Sciences, University of California, San
Diego, La Jolla, CA 92093, USA}

\author{C. H. Booth} \affiliation{Chemical Sciences Division, Lawrence
Berkeley National Laboratory, Berkeley, California 94720, USA}

\author{G. H. Kwei} \affiliation{Lawrence Livermore National
Laboratory, Livermore, California, 94550, USA} \affiliation{Los Alamos
National Laboratory, Los Alamos, New Mexico, 87545, USA}

\author{R. Chau} \affiliation{Lawrence Livermore National Laboratory,
Livermore, California, 94550, USA}

\author{M. B. Maple} \affiliation{Department of Physics and Institute
For Pure and Applied Physical Sciences, University of California, San
Diego, La Jolla, CA 92093, USA}

\date{preprint, March 29, 2002}

\begin{abstract}
The \ucupdx{} system exhibits non-Fermi liquid (NFL) behavior in 
thermodynamic and transport properties at low temperatures for Pd
concentrations $0.9 \lesssim x \lesssim 1.5$.  The local structure
around the U, Cu, and Pd atoms has been measured for 0 $\leq x \leq
1.5$ using the x-ray absorption fine structure (XAFS) technique in
order to quantify the effects of lattice disorder on the NFL
properties.  A model which allows a percentage of the Pd atoms to
occupy nominal Cu (16$e$) sites, $s$, was used to fit the Pd and Cu
$K$ edge and U $L_{\textrm{III}}$ edge data.  Pd/Cu site interchange was found
to occur in all samples ($x \neq 0$), reaching a minimum value of $s
\sim 0.17$ at $x = 0.7$ and increasing monotonically to $s \simeq 0.4$ at
$x=1.5$.  These data also determine an upper limit on the static disorder of 
the nearest neighbor U--Cu pairs.  A single-ion Kondo disorder model with a 
lattice-disorder origin of the distribution of $f$/conduction electron 
hybridization strengths within a tight-binding approach is used to calculate 
magnetic susceptibility.  The results indicate that the measured U--Cu
static disorder is not sufficient to explain the NFL behavior of the
magnetic susceptibility within  this variant of the Kondo disorder model,
suggesting either that other sources of Kondo disorder exist 
or that the Kondo disorder model is not applicable to \ucupdx.

\end{abstract}



\maketitle
%


\section{Introduction}

In $f$-electron compounds, strong electronic correlations lead to a
variety of interesting ground states.  Recently, there has been
growing interest in heavy fermion compounds that display non-Fermi
liquid (NFL) behavior in which the low temperature properties, such as
the electronic specific heat coefficient $C(T)/T \equiv \gamma$,
magnetic susceptibility $\chi(T)$, and electrical resistivity
$\rho(T)$, are characterized by logarithmic or power law temperature
dependences.  These characteristic temperature dependences are in
contrast to those of a Fermi liquid in which $\gamma \sim \chi \sim$
constant, and $\rho(T) \propto T^{2}$ at low
temperatures.\cite{Landau57} The NFL behavior of many of these Ce-,
Yb-, or U-based heavy fermion materials is often observed when long
range magnetic order is suppressed by the substitution of a nonmagnetic element
or by the application of pressure (for a review, see Refs. 
\onlinecite{Maple94,Schofield99,NFL_mat_ex,vonLohneysen99,Stewart01}).

There have been a number of theories proposed to account for non-Fermi
liquid behavior, although in many cases, several models adequately
describe the low temperature properties, making an unambiguous choice
of theories difficult.  In some systems, the NFL properties may be
associated with the proximity to a $T=0$ K second order phase transition
or quantum critical point (QCP).\cite{Hertz76,Millis93,Continentino94}
Since many materials that display non-Fermi liquid phenomena are
disordered alloys, various disorder-driven mechanisms have also been
proposed.  The ``Kondo disorder'' model
(KDM)\cite{Bernal95,Miranda96,Miranda97b} is essentially a
single-impurity model with a distribution of Kondo temperatures, and
might be considered as a disordered Fermi liquid model.  The KDM
utilizes a distribution of Kondo temperatures $T_{K}$ caused by a
distribution of $f$-electron/conduction electron exchange coupling strengths
$\mathcal{J}$, possibly induced by lattice disorder.  This model was
first put forth to explain the broad NMR linewidths and muon spin
rotation measurements on \xlow.\cite{Bernal95,MacLaughlin98}
Measurements of electrical resistivity, specific heat, and magnetic
susceptibility are also consistent with this
model.\cite{Andraka93,Chau96} Castro Neto {\em{et al.}} proposed a
model\cite{CastroNeto98,CastroNeto00} which describes the competition
between the RKKY and Kondo interactions in a disordered material
leading to the formation of magnetic clusters similar to a Griffiths'
phase.\cite{Griffiths69} This theory predicts that the physical
properties of the system at low temperatures are characterized by weak
power law behavior and is consistent with what is observed
experimentally in a number of NFL
materials.\cite{deAndrade98,Vollmer00} Recently, Miranda and
Dobrosavljevi\'{c} investigated a disordered Anderson lattice and
proposed that a Griffiths' phase could be obtained by a
disorder-driven metal-insulator transition
(MIT).\cite{Miranda01a,Miranda01b}

It is essential to determine the nature of the lattice disorder in
these disordered non-Fermi liquid systems, since it is a key
ingredient in many of the theories put forth and, therefore, may
provide valuable information regarding the underlying mechanism for
the NFL behavior.  The \ucupdx{} system was chosen for this
investigation because of its rich phase diagram and because one type
of disorder, Pd/Cu site interchange, in which Pd ions occupy nominal
Cu (16$e$) sites, was found to be relevant to the NFL properties at
low temperatures for UCu$_{4}$Pd.\cite{Booth98c}  Lattice disorder has
also been shown to be important from changes in the
lattice parameter and $\gamma$ in annealing studies of UCu$_4$Pd (Ref. 
\onlinecite{Weber01})
and the Cu-NQR lineshape in UCu$_{3.5}$Pd$_{1.5}$,\cite{Ambrosini99} and is 
implied from the 
glassy nature of the spin dynamics measured from the muon spin-lattice 
relaxation.\cite{MacLaughlin01}  The details of the low temperature magnetic 
phase diagram are much more complicated than originally believed.  In \ucupdx,
antiferromagnetic order is suppressed from $T_{N} = 15$ K in \ucu{},
to $T=0$ K at $x \sim 1$.\cite{Chau96,Andraka93} Although no frozen 
moments are observed from muon spin-relaxation, \cite{MacLaughlin01} 
glassy  behavior (spin
glass or superparamagnetism) is observed for $x = 1$ at $T_{f} \sim
0.2$ K, while no evidence for magnetic order down to 30 mK is found
for $x=1.25$.\cite{Scheidt98,Vollmer00} For $x=1.5$, spin glass
behavior reappears at $T_{f} \sim 0.1$ K which persists up to $x=2.2$
($T_{f} \sim 2 $ K), at which point the system enters into a mixed
phase region with two different crystal structures.  Therefore, the
\ucupdx{} system offers an opportunity to study the effects of
disorder on the relevant physical properties (i.e., magnetic order,
NFL behavior, etc.).  
In addition, it is still an open question as to
which model for non-Fermi liquid behavior is most appropriate for
this system.  A quantum critical point may exist given the
absence of magnetic order of any kind at $x=1.25$.  The various
Griffiths' phase models\cite{CastroNeto98,Miranda01a} may be
applicable since disorder is present in this system.\cite{Booth98c}
The Kondo disorder scenario may also be valid as many of the physical
properties are consistent with this 
model.\cite{Bernal95,MacLaughlin98,Chau96}

The main motivations for this study are: (1) to determine the amount
of lattice disorder in \ucupdx; (2) to calculate the distribution of
hybridization strengths $P(V)$ using the measured lattice disorder;
(3) to test the caveat of lattice disorder as a possible starting
point for a true microscopic model of the Kondo disorder model proposed in a previous 
study of Pd/Cu site interchange in UCu$_{4}$Pd;\cite{Booth98c} (4)
and, finally, to speculate on the applicability of models other than
the KDM to describe the NFL behavior of \ucupdx.  The x-ray absorption
fine-structure (XAFS) technique is a powerful tool for studying
disorder since it is a local structural probe that is atomic species
specific.  We will present a method for quantitatively determining the
amount of disorder around the U atoms in the \ucupdx{} system.  This
method can be applied to many systems in which it is helpful to
understand the nature of the disorder (or lack thereof) and its
relation to the physical properties.

The outline of the paper is as follows: experimental details and a
brief description of the XAFS technique are discussed in Sec. 
\ref{experiment}.  The theoretical background for the KDM along with
fits of this model to the magnetic susceptibility are presented in
Sec.  \ref{KDM}.  The experimental analysis methods, Pd and Cu $K$
edge and U $L_{\textrm{III}}$ edge XAFS data and results from
\ucupdx{} ($0 \leq x \leq 1.5$) samples are given in Sec. 
\ref{results}.  The effects of the measured lattice disorder on the
electronic properties of the \ucupdx{} system are considered in Sec. 
\ref{effects}.  A discussion of the results are presented in Sec. 
\ref{discussion} and conclusions are given in Sec.  \ref{conclusions}.

\section{Experimental Details}
\label{experiment}

Samples of \ucupdx{} were prepared by arc melting appropriate amounts
of the end members \ucu{} and \upd{} in an ultra high purity Ar atmosphere with a Zr
getter.  The parent compound \ucu{} was annealed in an evacuated
quartz ampoule at 900$^{\circ}$ C for 2 weeks, while no annealing was
performed on the compounds containing Pd.  All of the materials were
found to be single phase with the AuBe$_{5}$ crystal structure with no
observable impurity peaks according to x-ray diffraction measurements. 
A slightly Cu deficient sample, with composition UCu$_{4.95}$, was
also prepared in order to minimize Cu inclusions.\cite{Nakamura94}
However, the behavior of \ucu{} and UCu$_{4.95}$ were very similar and
only the data for the \ucu{} compound is reported except where
otherwise noted.  Magnetic susceptibility measurements were performed
in a commercial SQUID magnetometer in a magnetic field of 1 tesla at
temperatures between 1.8 K and 300 K.

X-ray absorption data were collected at the Stanford Synchrotron
Radiation Laboratory from the U $L_{\textrm{III}}$, Pd $K$, and Cu $K$
edges on BL 4-3 and BL 10-2 using a half-tuned double crystal Si(220)
monochromator with a slit height of 0.7 mm.  The samples were ground
to a fine powder in acetone, passed through a 30 micron sieve, and
brushed onto scotch tape.  Since \ucu{} oxidizes
rapidly,\cite{Sarma85} such powder samples were prepared in an Ar
glove box and sealed between layers of capton tape and shipped in a
quartz tube filled with Ar.  The capton-sealed \ucu{} and
UCu$_{4.95}$ samples were exposed to air for less than ten minutes
while being loaded into the cryostat.  Various numbers of layers were
stacked so that the edge step was approximately unity for each type of
edge.  Generally, 2-5 scans were collected for each sample for a
particular edge at temperatures between 3 K and 300 K with a
temperature deviation of less than 0.2 K.

\begin{figure}
\includegraphics[width=3.4in]{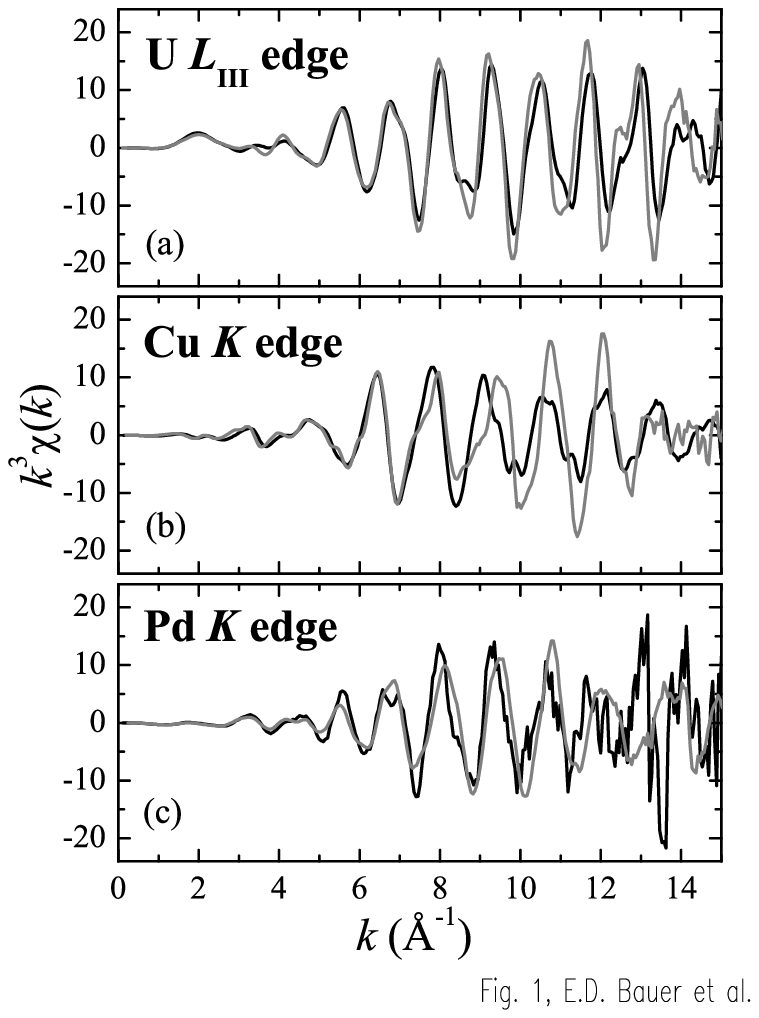}
\caption{Examples of the $k^3\chi(k)$ XAFS data for the $x$=0.3 (dark)
and the $x$=1.5 (light) samples from the (a) U $L_{\textrm{III}}$, (b)
Cu $K$, and (c) Pd $K$ edges.  Note the relatively poor quality of the
Pd $K$ edge data for $x$=0.3.  The data quality of other samples is
similar to that of the $x$=1.5 sample.  }
\label{kspace}
\end{figure}

The absorption data were reduced using standard procedures outlined in
Refs.  \onlinecite{Hayes82} and \onlinecite{Li95b}, including 
fitting an embedded-atom absorption function $\mu_{0}(E)$ with a
5-7-knot cubic spline function with a maximum photoelectron wavevector $k$
of 15 \AA$^{-1}$.  The XAFS function is then defined as
$\mu(k)/\mu_{0}(k) -1$, where $\mu$ is the absorption coefficient,
\(k=\sqrt{\frac{2m_{e}}{\hbar^{2}}(E - E_{0})}\) is the photoelectron
wavevector, $m_{e}$ is the electron rest mass, and $E_{0}$ is the
absorption edge threshold energy, which is defined arbitrarily to be
the half-height of the edge and allowed to vary in the fits.  Examples
of the $k$-space data for each of the absorption edges are shown in
Fig.  \ref{kspace}.

The XAFS technique provides local structural information about an
absorbing atom.  The oscillation above an absorption edge, which are
due to the interference between the outgoing and backscattered
photoelectron waves, can be Fourier transformed and subsequently fit
in order to determine the following quantities: $N_{i}$, the number of
atoms in the $i^{{\textnormal{th}}}$ shell at a radius $R_{i}$ from
the absorbing atom; $\sigma_{i}$, the spread in bond lengths at
$R_{i}$ (also called the Debye-Waller factor); and $S_{0}^{2}$, an
overall amplitude reduction factor that accounts for inelastic losses. 
Theoretical predictions for $S_{0}^{2}$ generally do not compare well
with experiment; therefore, $N_{i}$ is held fixed whenever possible. 
In this study, we fit the Fourier Transform (FT) of the XAFS function
$k^{3}\chi(k)$ to the theoretical backscattering and phase functions
calculated by FEFF7\cite{FEFF6} (more details can be found in Ref. 
\onlinecite{Li95b}).  The absolute errors in $\sigma$ are estimated to
be 5\% for nearest neighbor bonds and 10\% for further neighbor bonds. 
The absolute errors in $R$ are estimated to be 0.01 \AA{} for nearest
neighbor bonds and 0.02 \AA{} for further neighbor bonds.\cite{Li95b}
These absolute error estimates are similar to (though always larger
than) those obtained by other methods such as a Monte Carlo
method\cite{Lawrence01} or the relative errors estimated from the
number of scans taken at a given temperature for all the measurements
report here.

As an additional check on the sample purity of the \ucupdx{} samples,
a contribution to the XAFS of a number of impurities were considered
such as Cu, Pd, CuPd, Cu$_{3}$Pd, as well as various oxides. 
Simulations of the XAFS spectra expected from nearly all of these
impurity compounds indicated that the tested impurities would produce
peaks that were not observed in the \ucupdx{} data and therefore these
impurities could only be present at levels of less than a few percent. 
The impurity level of compounds with similar crystal structures and
lattice parameters, such as free Cu (which was reported to exist in
UCu$_{4.35}$Pd$_{0.65}$ (Ref.  \onlinecite{Chau98})), could be as high
as 5\%.

\section{Background}
\label{KDM}

\subsection{Magnetic susceptibility and logarithmic behavior}

\begin{figure}
\includegraphics[width=3.3in]{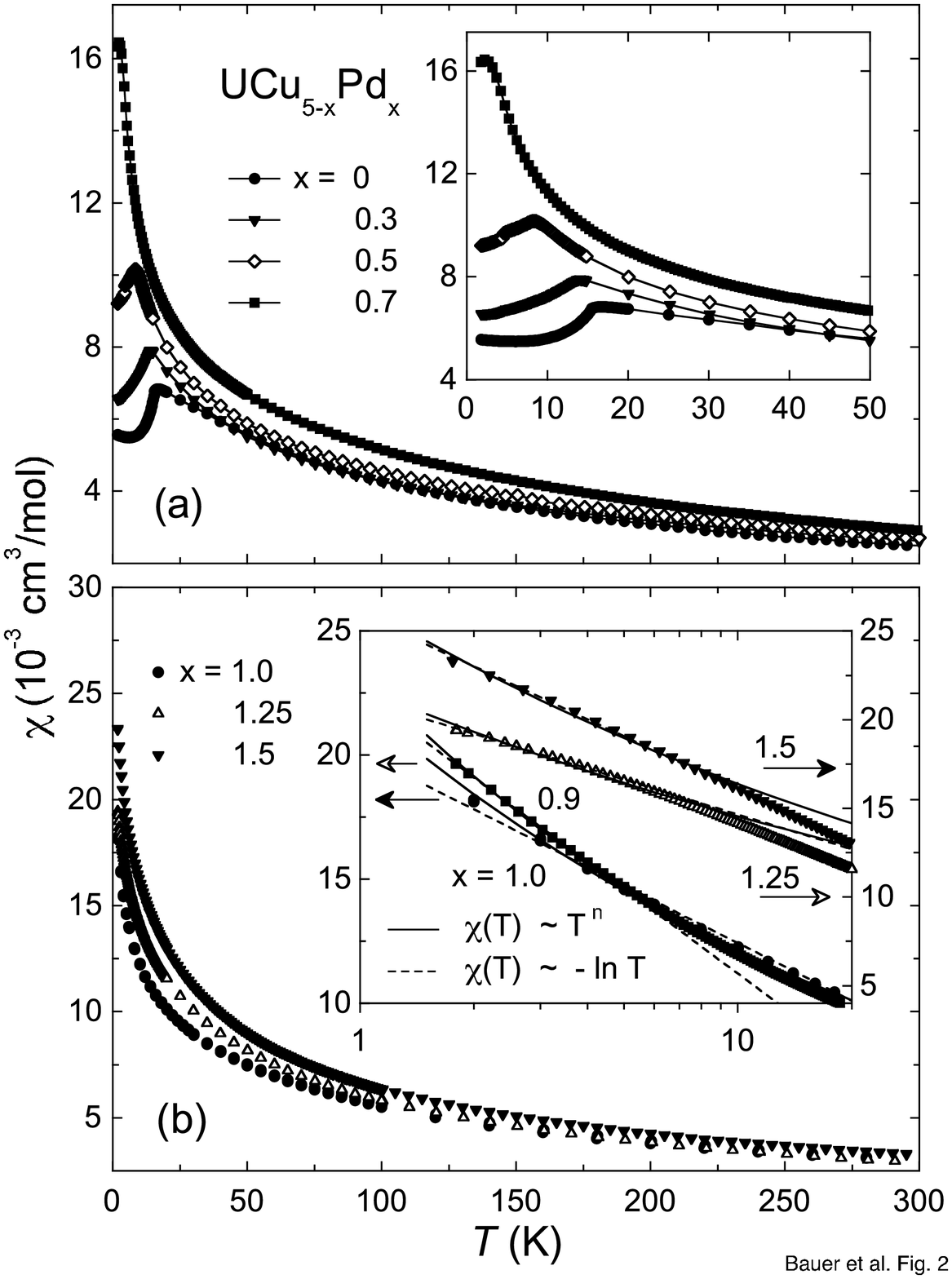}
\caption{Panel a): Magnetic susceptibility $\chi \equiv M/H$ vs
temperature $T$ of \ucupdx{} compounds with Pd concentrations in the
range $0 \leq x \leq 0.7$ in a magnetic field  $H$=1 T. Inset:
$\chi(T)$ below 50 K. Panel b): Magnetic susceptibility $\chi(T)$ in
$H$=1 T for \ucupdx{} compounds with Pd concentrations in the range
$1.0 \leq x \leq 1.5$.  Inset: $\chi(T)$ at low temperature for Pd
concentrations between $x=0.9$ and 1.5.  The lines are fits of the
expressions \(\chi(T) = \chi_{0}- c\, ln\,T\) (dashed lines) and
\(\chi(T) \propto T^{n} \) (solid lines) to the $\chi(T)$ data.}
\label{chi}
\end{figure}

Measurements of magnetic susceptibility were performed in order to
track the decrease of the N{\'{e}}el temperature with increasing Pd
concentration and to explore the non-Fermi liquid behavior observed
previously.\cite{Andraka93,Chau96} The magnetic susceptibility $\chi
\equiv M/H$ vs temperature $T$ in a magnetic field of $H= 1$ T for
various \ucupdx{} compounds is shown in Fig.  \ref{chi}.  A cusp or
peak of the $\chi(T)$ curves is observed for $0 \leq x \leq 0.7$,
indicating the onset of antiferromagnetic order at $T_{N}$ = 15.5 K,
14.1 K, 8.4 K, and 2.3 K for $x$=0, 0.3, 0.5, and 0.7, respectively,
in agreement with previous results.\cite{Andraka93,Chau96} For $0.9
\leq x \leq 1.5$, the $\chi(T)$ data increases with decreasing
temperature at low $T$ and can be fitted by the expressions \(\chi(T)
= \chi_{0} - c \ln T\) where $\chi_{0}$ and $c$ are constants, or
\(\chi(T) = - c^{\prime} T^{n} \) where $c^{\prime}$ is a constant and
$n$ ranges between -0.2 and -0.3, as shown in the inset of Fig. 
\ref{chi}.  Similar power law behavior has been previously observed in \ucupdx{}
(Ref.  \onlinecite{Vollmer00}) and other disordered NFL systems such
as Y$_{1-x}$U$_{x}$Pd$_{3}$ and
Th$_{1-x}$U$_{x}$Pd$_{2}$Al$_{3}$.\cite{deAndrade98} 
At higher
temperatures above $\sim 200$ K, the magnetic susceptibility of the
\ucupdx{} compounds follows a Curie-Weiss law:
\begin{equation*}
	\chi\, =\, C/(T-\theta_{CW})
\end{equation*}
where $C=N_{A}\mu_{eff}^{2}/3k_{B}$, $\theta_{CW}$ is the Curie-Weiss
temperature and $\mu_{eff}$ is the effective moment in Bohr magnetons. 
The effective moment $\mu_{eff}=3.0 - 3.4$ $\mu_{B}$ is smaller than
what would be expected for a 5$f^{2}$ ($\mu_{eff}=3.58$ $\mu_{B}$) or
5$f^{3}$ ($\mu_{eff}=3.62$ $\mu_{B}$) configuration, indicating the
presence of crystalline electric field effects and/or the Kondo
effect.  The Curie-Weiss temperatures $\theta_{CW}$ are negative for
all concentrations studied and increase roughly linearly from -260 K
to -100 K from $x=0.3$ to $x=1.5$ (results listed in Table \ref{mag}).

\begin{table*}	
\caption{Magnetic properties of \ucupdx.  The N{\'{e}}el temperature
$T_{N}$ is obtained from the change in slope of  the $\chi(T)$ curves. 
The values of the effective moment $\mu_{eff}$ and Curie-Weiss
temperature $\theta_{CW}$ are extracted from fits to a Curie-Weiss law
of the high temperature magnetic susceptibility.  The parameters
$\chi_{0}$ and $c$ are determined from fits of the expression
\(\chi(T) = \chi_{0} - c\, ln\,T\) to the $\chi(T)$ data.  The
exponent $n$ is determined from fits of the expression \(\chi(T)
\propto T^{n} \) to the $\chi(T)$ data.  The values of the static
disorder necessary to fit the magnetic susceptibility $\sigma_{KDM}$
are determined from fits of the KDM to the $\chi(T)$ data, and include
the Cu and Pd site distributions involving the Pd concentration $x$
and the site interchange $s$, as discussed in Sec.  \ref{Pdedge}.}
\begin{ruledtabular}
\begin{tabular}{ldccdcccd}
x & \multicolumn{1}{c}{$T_{N}$ (K)} & $T_{f}$ (K) &$\mu_{eff}$ ($\mu_{B}$) & 
\multicolumn{1}{c}{$\theta_{CW}$ (K)} & 
 $\chi_{0}$ (10$^{-3}$ cm$^{3}$/mol)  & $c$ (10$^{-3}$ cm$^{3}$/mol 
 ln K)   & $n$ & \multicolumn{1}{c}{$\sigma_{KDM}^{2}$ (\AA$^{2}$)} \\\colrule  
0 & 15.5 &  & 3.1 & -186 &  &    & & \\
0.3 & 14.1 & & 3.4 & -259 &  &   & & \\
0.5 & 8.4 & & 3.3 & -198 &  &    & & \\
0.7 & 2.3 & & 3.3 & -162 &  &  &   &  \\
0.9 & 0.5\footnote{Ref. \onlinecite{Korner00}} 
         &  & 3.2 & -134 & 23 & 4.9  &  -0.29 & 0.0038\\
1.0 & 0.2\footnote{Ref. \onlinecite{Korner00}}  & 0.2\footnote{Ref. 
\onlinecite{Korner00,Vollmer00,Scheidt98}}
         & 3.2 & -139 & 20 & 3.3  & -0.26 & 0.0030\\
1.25 &  &  & 3.0 & -97 & 21 & 2.8   & -0.18 & 0.0028 \\
1.5 &  & 0.1\footnote{Ref. \onlinecite{Vollmer00,Scheidt98}}
         & 3.2 & -107 & 26 & 4.3   & -0.21 &0.0028 \\
\end{tabular}
\end{ruledtabular}
\label{mag}
\end{table*}

\subsection{Lattice disorder and the KDM}

In the Kondo disorder model, a logarithmic divergence at low temperatures in the magnetic
susceptibility $\chi(T)$ and specific heat divided by temperature
$C(T)/T$ are produced when a sufficient amount of weight in the
distribution of Kondo temperatures $P(T_{K}$) exists at
low-$T_{K}$.\cite{Bernal95} The P($T_{K}$) distribution is determined
by introducing a spread of coupling strengths, $\mathcal{J}N(E_{F})$,
into the expression for the Kondo temperature
\begin{equation}
	k_{\textrm B}T_{K} = E_{F} e^{-1/\mathcal{J}N(E_{F})}
\label{TK}
\end{equation}
where $k_{\textrm B}$ is Boltzmann's constant, $\mathcal{J}$ is the
local moment/conduction electron exchange parameter and $N(E_{F})$ is
the  electronic  density of states of the host material at the Fermi level $E_F$. 
The exchange parameter is given by:
\begin{equation}
   \mathcal{J} = - \frac{V_{tot}^{2}}{|\epsilon_{f}|},
\label{J}
\end{equation}
where $V_{tot}$ is the matrix element which admixes the local moment
and conduction electron states, and $\epsilon_{f}$ is the $f$-ion
energy below the Fermi level.  

The variant of the Kondo disorder model considered here utilizes a
tight binding approach to compute the hybridization strengths $V$
between the conduction electron states and the $f$-states.   Since the
$d$-electron contribution is known to dominate the conduction band in
isostructural YbAgCu$_{4}$,\cite{Monachesi96} we only consider the
$V_{fd}$ matrix element.  The tight binding formalism of Harrison and
Straub for each pair of atoms $U$ ($f$ ion) and $X$ ($d$ electrons in
the conduction band) then gives:\cite{Harrison87}
\begin{equation}
V_{fd} =
\frac{\eta_{fd}\hbar^{2}}{m_{e}}\frac{\sqrt{(r_{Uf}^{5}r_{Xd}^{3})}}
{R_{U-X}^{6}},
\label{vfd}
\end{equation}
where $r_{Xl}$ is the radius of the electronic shell of atom $X$ with
angular momentum $l$ (=$d$ or $f$) (tabulated in Ref. 
\onlinecite{Straub85}), $R_{U-X}$ is the bond length between the $U$
and $X$ atoms, and $\eta_{fd}$ is a factor which only depends on the
$l$'s and bond symmetry (see Appendix B of Ref. 
\onlinecite{Harrison87}).  The total coupling energy is obtained by
summing over all atoms pairs
\begin{equation}
	V_{tot}=\sum_{U,X Pairs} V_{fd}.
\label{vtot}
\end{equation}
Within this approach, the coupling energies are often larger than
experimentally observed energies by a factor of two.\cite{Harrison87}
This limitation does not play a major role in what follows because 
 these energies are scaled by $N(E_F)/\epsilon_f$, 
which is used as a fitting parameter.  Therefore,  we
only need consider relative changes in $V_{fd}$.  The sum in Eq. 
\ref{vtot} converges rapidly since the contribution from more distant
neighbors is small, i.e., $V_{fd}$($R_{\rm{U-Cu}}=4.59$ \AA{})
$\sim$ 7\% of $V_{fd}$($R_{\rm{U-Cu}}=2.93$ \AA{}); therefore, we
assume that only the first two nearest neighbor shells contribute to
the hybridization.

We include two types of contributions to the distribution of
hybridization strengths due to lattice disorder.  One effect is that
of Pd/Cu site interchange, which changes the species of atoms used in
the $r_{Xd}$ parameters and will result in different hybridization
strengths, i.e., the hybridization will be larger for a U atom with
more Pd neighbors than one with more Cu neighbors.  The distribution
of hybridization strengths due to the different $r_{Xd}$ parameters is
therefore discrete and modeled by a binomial distribution.  The other
contribution to $P(V)$ is due to continuous lattice disorder, which is
included as a Gaussian distribution of bond lengths $R_{U-X}$ with
width $\sigma_{static}$. The distribution of Kondo temperatures is calculated with the
use of the distribution of hybridization strengths $P(V)$ in Eq. 
\ref{J}.  With increasing continuous disorder, $\sigma_{static}$, the
initial discrete $P(T_{K}$) is quickly washed out even for small
amounts of disorder $\sigma_{static} \sim 0.005$ \AA{}, and broadens
considerably with more weight shifting to lower $T_{K}$.\cite{Booth01}
Sources of 
Kondo disorder other than a bond length distribution in the hybridization energy
such as a distribution of $f-$ ion energies or density of 
states $N(E_{F}$ are not included in this model.

\subsection{Fits of the KDM to the magnetic susceptibility}

\begin{figure}
\includegraphics[width=3.3in]{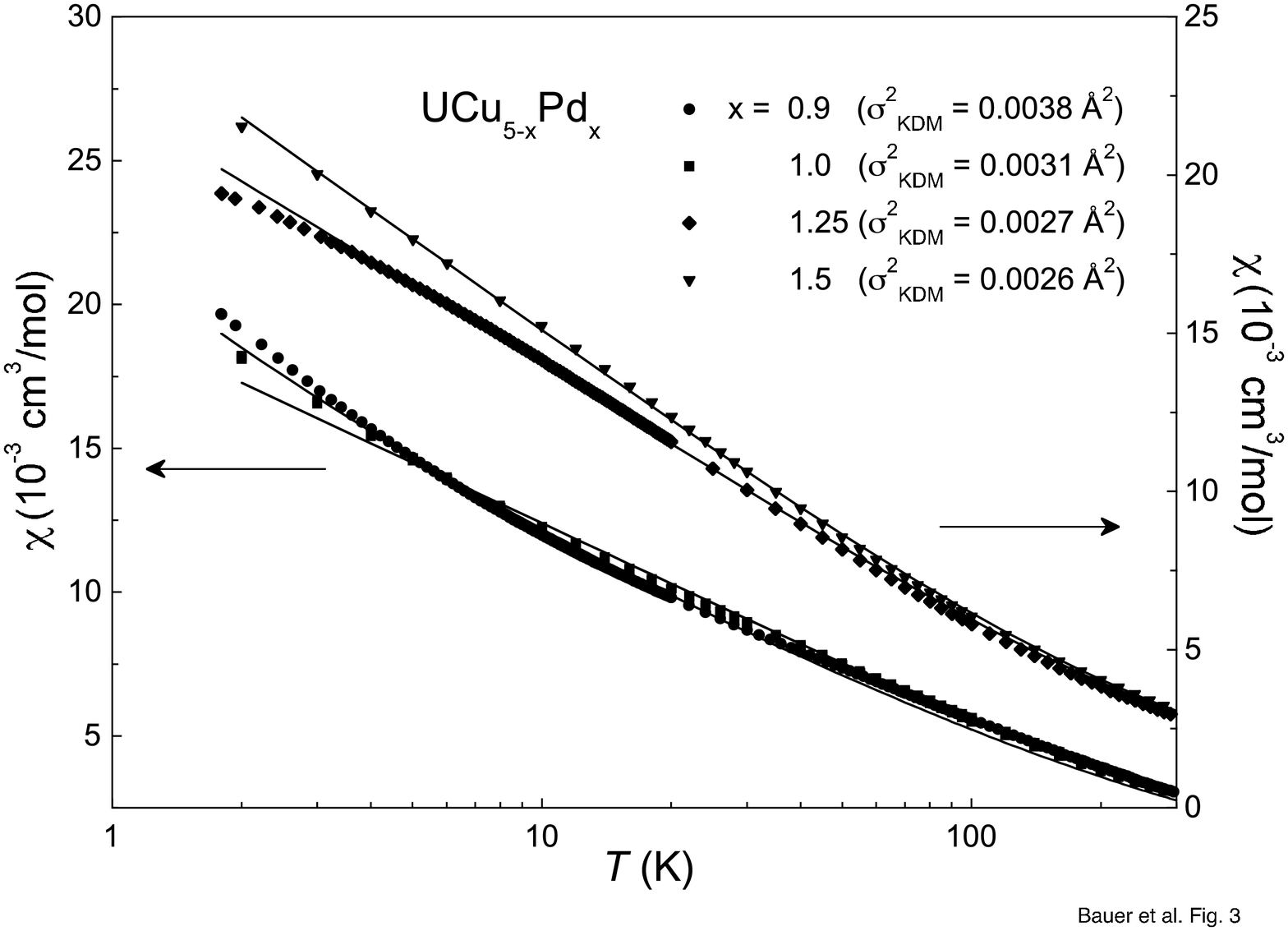}
\caption{Fits of the Kondo disorder model to the
$\chi(T)$ data of \ucupdx{} for Pd concentrations between $x=0.9$ and
$x=1.5$.}
\label{NFLchi}
\end{figure}

The magnetic susceptibility was calculated within the KDM by
convolving $P(T_{K}$) with the theoretical $\chi(T,T_{K})$ proposed by
Rajan\cite{Rajan83} for a $J$=3/2 Kondo impurity.  Three parameters
were used to fit the model to the data: continuous static disorder
$\sigma_{KDM}$, $E_{F}$, and $N(E_{F})/\epsilon_{f}$.  The latter two
variables were found to be roughly constant for all concentrations at
values of 1.36 eV (similar to the value of $E_{F}$ of 1.41 eV used in
Ref.  \onlinecite{Bernal95} (for $x=1$)) and 0.157 eV$^{-2}$,
respectively.  Least-squares fits to the $\chi(T)$ data are shown in
Fig.  \ref{NFLchi} (results collected in Table \ref{mag}).  The fit
quality is  excellent, especially  considering that the susceptibility was derived
from a microscopic model with only three adjustable parameters and the
simplifying assumptions of the tight binding model and single-ion
behavior.  Note that, for instance, a value of $\sigma_{KDM}^{2} \sim
0.0031$ \AA$^{2}$ is necessary to produce enough weight at low $T_{K}$
to generate the logarithmic divergence in $\chi(T)$ at low
temperatures for the $x$=1.0 sample.  The amount of static disorder used in the KDM fits
decreases with increasing Pd concentration.

\section{XAFS Results and Analysis}
\label{results}

\subsection{Experimental Analysis Procedure}

\begin{figure}
\includegraphics[width=2.5in,clip=true]{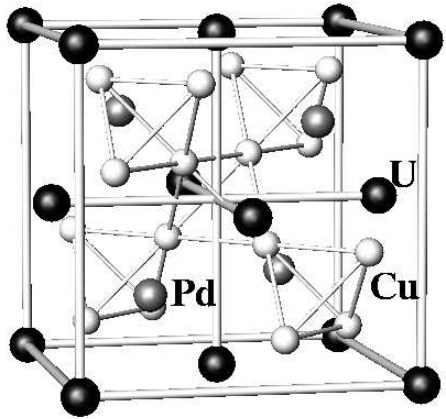}
\caption{Crystal structure of ordered UCu$_{4}$Pd.  U atoms occupy the
$4a$ sites (black spheres), Pd atoms occupy the $4c$ sites (shaded
spheres), and the Cu atoms occupy the $16e$ sites (white spheres).}
\label{xtal}
\end{figure}

Before discussing the detailed analysis procedure, it is instructive
to describe the average and local structure of the \ucupdx{} system. 
This system crystallizes in the face centered cubic AuBe$_{5}$
structure (C15$b$) for $x < 2.3$.  We consider for this discussion an
ordered UCu$_{4}$Pd structure in which the U and Pd atoms ($4a$ and
$4c$ sites, respectively) form interpenetrating FCC lattices and the
Cu atoms form a network of vertex-sharing tetrahedra along the body
diagonal ($16e$ sites) as shown in Fig.  \ref{xtal}.

\begin{figure}
\includegraphics[width=2.3in]{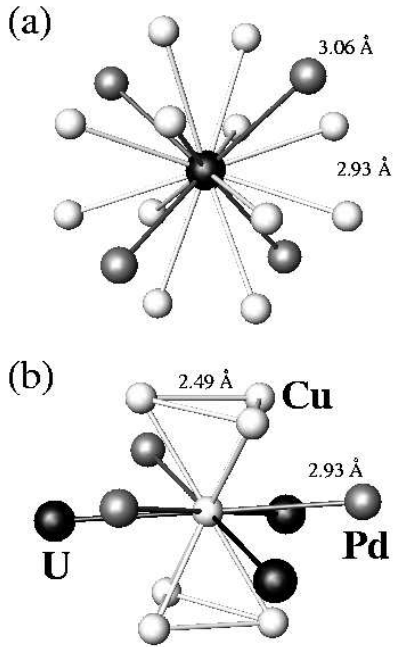}
\caption{Local structure in ordered UCu$_{4}$Pd.  Panel a) shows the
local structure about the U atoms (the Pd local structure is obtained
by switching all of the U and Pd atoms).  Panel b) shows the local
structure about the Cu atoms.}
\label{local}
\end{figure}

The local environments of the U  and Cu atoms  are shown in Fig. 
\ref{local}a and Fig.  \ref{local}b, respectively.  The U and Pd
environments are very similar since the U (Pd) atoms have 12 Cu
neighbors at 2.93 \AA{} and 4 Pd (U) neighbors at 3.06 \AA. The local
structure about the Cu atoms is different due to their tetrahedral
arrangement with 6 Cu neighbors at 2.50 \AA{} and both 3 U
and Pd neighbors at 2.93 \AA. The site interchange causes part of the
Pd XAFS spectrum to be due to Pd atoms on 4$c$ sites and part to be
due to Pd atoms on 16$e$ sites.  The main difference between these two
spectra will be due to the short Pd-Cu pair at $\sim$2.5 \AA{} arising from
the Pd atoms on $16e$ sites.

In order to understand the effects of lattice disorder on the
electronic properties in the \ucupdx{} system, one must be able to
measure the detailed local structure environment around the uranium
atoms.  Determining this local structure arrangement is a natural
measurement for the XAFS technique since this technique is atomic
specific, allowing the uranium atoms to be singled out.  In addition,
the Debye-Waller factors have been shown to be meaningful in an
absolute sense, as was demonstrated for cubic materials in
general,\cite{Li95b} and for isostructural Yb$X$Cu$_{4}$ ($X$ = Ag,
Cd, In, Tl) in particular.\cite{Lawrence01} However, the presence of
Pd/Cu site interchange makes the analysis procedure more complex
because the species of atom in each of the first two shells cannot
necessarily be well determined in the fits to the uranium data alone.

Given the difficulty of determining the detailed local structure
around the uranium atoms in \ucupdx, we have developed the following
approach.  First, we use the Pd $K$ edge XAFS data to determine the
fraction of Pd atoms on nominal Cu ($16e$) sites $s \equiv N$(Pd on
$16e$ sites)/$N$(Total Pd).  If, for instance, the Pd atoms are
randomly distributed on all 16$e$ and $4c$ sites for UCu$_4$Pd, the
value of $s$ would be 0.8 (4/5 of all possible sites are 16$e$ sites). 
Once values of $s$ (and errors on $s$) are determined, these can be
held fixed in fits to the U $L_{\textrm{III}}$ edge data, and a
reliable value of the square of the Debye-Waller factor of the
nearest-neighbor U--Cu pair $\sigma^2_{\rm U-Cu}$ can be determined. 
With these values of $\sigma^2_{\rm U-Cu}$ as a function of
temperature, one may then estimate the amount of static (i.e.,
non-thermal) pair distance disorder by fitting the U edge data to a
correlated-Debye model with a temperature-independent
offset:\cite{Crozier88}
\begin{equation}
	\sigma_{meas}^{2}(T,\theta_{cD}) = \sigma_{static}^{2} + F(T,
	\theta_{cD}).
\label{corr_debye}
\end{equation}
The temperature-dependent part of the Debye-Waller factor $F(T,
\theta_{cD}$) is given by the correlated Debye model
\begin{equation*}
F(T, \theta_{cD}) =\frac{\hbar}{2 \mu} \int \rho_{j}(\omega)
coth\left(\frac{\hbar \omega}{2 k_{B} T} \right)
\frac{d\omega}{\omega}
\end{equation*}
where $\mu$ is the reduced mass, $\theta_{cD}$ is the correlated Debye
temperature, and the phonon density of states at position $R_{j}$
is\cite{Beni76}
\begin{equation}
\rho_{j} = \frac{3 \omega^{2}}{\omega_{D}^{3}} \left[ 1 - 
\frac{sin(\omega R_{j}/c)}{\omega R_{j}/c}\right]
\label{dos}
\end{equation}
in which $\omega_{D}$ is the usual Debye frequency and $c=
\omega_{D}/k_{D}$.  The expression in brackets of Eq.  \ref{dos}
takes into account the correlated motion of the atom pair.  These
values of the static disorder $\sigma^2_{\rm static}$ can then be
compared to those ($\sigma_{KDM}$) necessary for the KDM to describe
the magnetic susceptibility data.  Debye-Waller factors for the
neighbors further than the nearest neighbor U--Cu pairs are judged to
be less reliable for the purpose of determining $\sigma^2_{\rm
static}$, either because the fraction of the amplitude of the peak in
the Fourier transform of the XAFS data due to the pair of interest is
less that 50\% of the peak height (as is the case for the
next-neighbor U-Pd/Cu pairs at about 3.1 \AA) or simply because the
correlated-Debye model does not work very well for further
neighbors.\cite{Li95b}

\subsection{Pd $K$ edge}
\label{Pdedge}

The FT of $k^{3}\chi(k)$ of the Pd $K$ edge of \ucupdx{}  for various Pd
concentrations is shown in Fig.  \ref{pd}.  Hereafter,
``Pd$^{\prime}\,$'' denotes Pd atoms on 16$e$ sites, ``Pd'' denotes Pd
atoms on 4$c$ sites, ``Cu$^{\prime}\,$'' denotes Cu atoms on 4$c$
sites, and ``Cu'' denotes Cu atoms on $16e$ sites.  The main peak at
$\sim 2.4$ \AA{} (peaks in the XAFS transforms are shifted from the
real bond lengths due to the phase shifts of the absorbing and
backscattering atoms) arises mainly from the large number of Pd-Cu
bonds at 2.93 \AA{} with additional contributions (shoulders) from the
Pd$^{\prime}$-Cu/Pd$^{\prime}$ bonds at 2.50 \AA{} and Pd-U bonds at
3.06 \AA{} with appropriate weights governed by the amount of site
interchange $s$.  The main peak at $\sim 2.4$ \AA{} shifts to lower
$r$ with increasing Pd concentration, indicating more of the $16e$
sites are being occupied by Pd (a situation which must occur for
$x>1$).  The model used to fit these data is extended over the basic
model used in Ref.  \onlinecite{Booth98c} by including all Pd bonds,
both on the 4$c$ sites and the 16$e$ sites, out to pair distances of
4.59 \AA. These fits are tightly constrained in order to reduce the
number of fit parameters to be well within Stern's rule.\cite{Stern93}
For instance, amplitudes for all the peaks are set by only three
parameters: $S_0^2$, $x$, and $s$, with a single $S_0^2$ used for all
data at a given edge and $x$ fixed to the nominal value.  In addition,
pair distances that are nominally the same, such as $R$(Pd--Cu) at
2.93 \AA{} in the nominal structure and
$R$({Pd$^{\prime}$--Cu$^{\prime}$) that occurs from the site
interchange, are held equal in the fits.  Different atom pairs with
equal pair distances also have the ratio of the Debye-Waller factors
$\sigma^2$ held to the ratio of their reduced masses, although this
does not account for differences in the correlated Debye temperatures
of the various bonds.  Finally, a single $S_0^2$ and threshold energy
shift $\Delta E_0$ are used for all bonds and all temperatures.  With
these constraints we include 15 atom pairs in the fits while only
requiring 10 fit parameters from 1.5-4.5 \AA, using a $k$-space window
of 2.5-15 \AA$^{-1}$ (Gaussian narrowed by 0.3 \AA$^{-1}$).  The value
of S$_{0}^{2}$ for the Pd $K$ edge data is determined to be 0.85
$\pm$ 0.04.  Representative fits for various Pd concentrations are
displayed in Fig.  \ref{pd} and summarized in Table \ref{pdfitstable}. 
The quality of the fits is quite good with $R \sim 3-10$\% for all Pd
concentrations measured and the bond lengths are consistent with
diffraction results.

\begin{figure}
\includegraphics[width=3.3in]{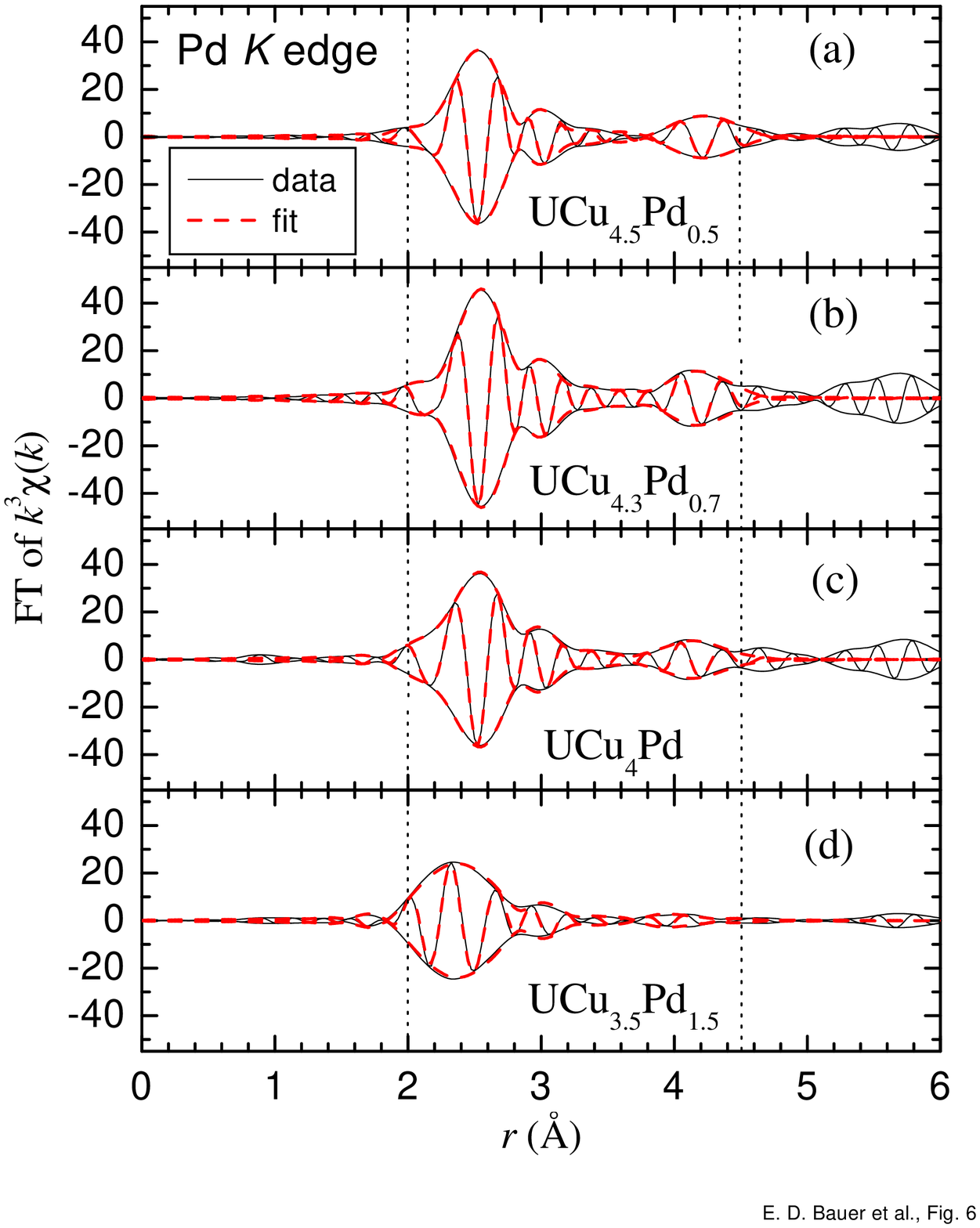}
\caption{Fourier transforms of $k^{3}\chi(k)$ of the Pd $K$ edge data from
selected \ucupdx{} samples.  The outer envelope is
the amplitude and the oscillating inner line is the real part of the
complex transform.  Solid lines are data for a)
$x=0.5$ (20 K), b) $x=0.7$ (15 K), c) $x=1.0$ (3.3 K), and d) $x=1.5$ (15 K)
along with fits
(dashed lines) as described in the text.  These and all subsequent 
transforms are from
$k$=2.5-15 \AA$^{-1}$ and Gaussian narrowed by 0.3 \AA$^{-1}$.  The
fit range is from 1.5 to 4.5 \AA{} (dotted lines).}
\label{pd}
\end{figure}

The amount of site interchange $s$ vs $x$ is shown in Fig. 
\ref{svsx}a along with the nominal values of $s$, in which the $4c$
(nominal Pd) sites are filled with Pd up to $x=1$ (i.e., $s$ = 0), and
thereafter the Pd atoms begin occupying the $16e$ (nominal Cu) sites
(i.e., $s> 0$).  These
results indicate that site interchange exists even for low Pd
concentrations (results listed in Table \ref{svsxtable}).  In a previous XAFS 
study of UCu$_{4}$Pd, the amount of Pd/Cu site interchange was found to be 
$s=0.24(3)$.\cite{Booth98c}  In  the present investigation, 
the data have been analyzed using a 
more sophisticated model of the local Pd environment (i. e. including further neighbors 
with site interchange out to 
4.6 \AA{} instead of 3.4 \AA\cite{Booth98c}) to obtain $s=0.32(5)$.  
While there is a small discrepancy between the two 
models, the values are consistent with each other within the error bars. 
Differences in the value of $S_{0}^{2}$ between the two studies may account for this discrepancy. 
In the present study, 
the value of the XAFS scale factor $S_{0}^{2}=0.85$ is based upon a number of Pd concentrations and therefore 
is judged to be more reliable.   

\begin{figure}
\includegraphics[width=3.3in]{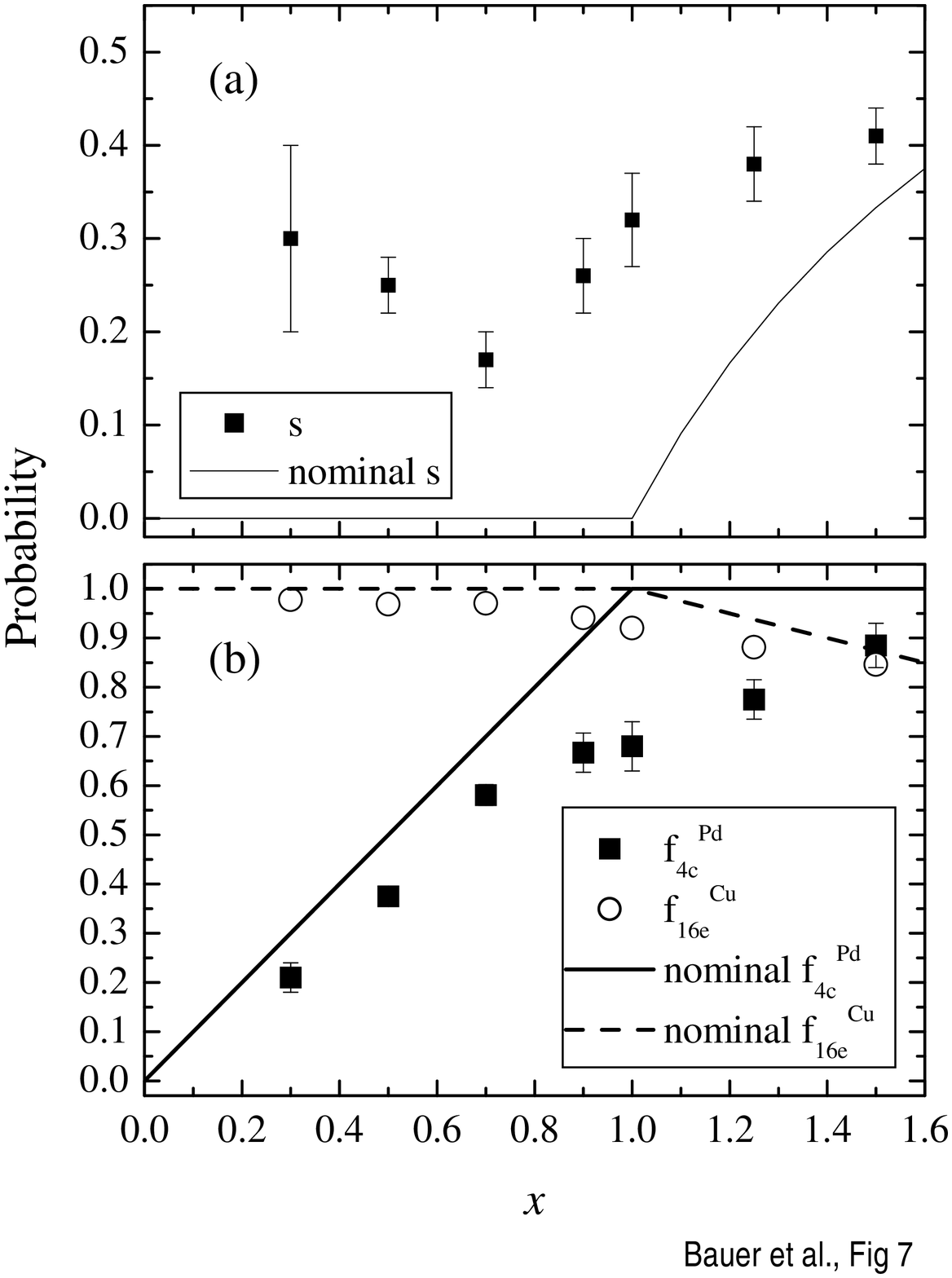}
\caption{Panel a) shows the amount of Pd/Cu site interchange $s$ vs Pd
concentration $x$ in \ucupdx{} determined from fits of the Pd $K$ edge
data to the site interchange model.  The nominal $s$ is based on a simple
model in which the Pd atoms first fill the $4c$ sites for $x<1$ and
thereafter fill the $16e$ sites.  Panel b) gives the occupancy fractions
$f_{4c}^{Pd}$ (fraction of Pd on $4c$ sites) and $f_{16e}^{Cu}$
(fraction of Cu on $16e$ sites) vs $x$.  The nominal occupancy factors
$f_{4c}^{Pd}$ and $f_{16e}^{Cu}$ are indicated by the solid and dashed
lines, respectively.}
\label{svsx}
\end{figure}


\begin{table*}
\caption{Fit results of the site interchange model to some of the Pd $K$ edge
data from \ucupdx{}.  $N$--nominal number of neighbors in ordered
UCu$_{4}$Pd; $A$--peak amplitude; $\sigma^{2}$--square of Debye-Waller
factor; $R$--bond length; $f^{Cu}_{16e}$--16$e$ site occupancy of Cu;
$f^{Pd}_{4c}$--4$c$ site occupancy of Pd.  The site-interchange ions
are denoted with a prime, e.g., Pd$^{\prime}$ is a Pd ion on a nominal
Cu ($16e$) site.  The amplitudes of the peaks $A$ are fixed by the
amount of Pd/Cu site interchange $s$ and the site occupancy factors
$f^{Cu}_{16e}$ and $f^{Pd}_{4c}$.  Atom pairs in parentheses are 
completely constrained by the pair above them in the table.
Debye-Waller factors for pairs with similar bond lengths are also constrained.  See Sec. \ref{Pdedge} for details.
$S_{0}^{2}$ is determined to be 0.85 $\pm$0.04.
The errors in $\sigma$ are estimated to be 5\% for nearest neighbor
bonds and 10\% for further neighbor bonds.  The errors in $R$ are
estimated to be 0.01 \AA{} for nearest neighbor bonds and 0.02 \AA{}
for further neighbor bonds.}
\label{pdfitstable}
\begin{ruledtabular}
\begin{tabular*}{\hsize}{lcddddddddd}
 \multicolumn{2}{l}{Pd Concentration $x$, $T$} &\multicolumn{3}{c}{0.7, 15 K}  & 
\multicolumn{3}{c}{1.0, 3.3 K} & \multicolumn{3}{c}{1.25, 15 K}\\
 Bond  &  $A$ constraint  & \multicolumn{1}{c}{$A$} &  
 \multicolumn{1}{c}{$\sigma^{2}$ (\AA$^{2}$)} & \multicolumn{1}{c}{$R$ (\AA)} &
\multicolumn{1}{c}{$A$} &  \multicolumn{1}{c}{$\sigma^{2}$ (\AA$^{2}$)} & 
\multicolumn{1}{c}{$R$ (\AA)} & \multicolumn{1}{c}{$A$} 
 & \multicolumn{1}{c}{$\sigma^{2}$ (\AA$^{2}$)} 
 & \multicolumn{1}{c}{$R$ (\AA)} \\\hline

Pd$^{\prime}$--Cu &  6$S^{2}_{0}\,s\,f^{Cu}_{16e}$ & 0.84 & 0.0029 & 
2.55 & 1.50 &
0.0028 & 2.56  & 1.71 & 0.0025 & 2.55  \\
(Pd$^{\prime}$--Pd$^{\prime}$) & 6$S^{2}_{0}\,s\,(1-f^{Cu}_{16e}$)  & & & & &\\ 

Pd$^{\prime}$--Cu$^{\prime}$  &  3$S^{2}_{0}\,s\,(1-f^{Pd}_{4c}$) & 
0.18 & 0.0030 & 
2.91 & 0.26 & 0.0025 & 2.90 & 0.22 & 0.0025 & 2.88   \\
(Pd$^{\prime}$--Pd) &  3$S^{2}_{0}\,s\,f^{Pd}_{4c}$  & & &\\ 
(Pd$^{\prime}$--U) & 3$S^{2}_{0}\,s$  & & &\\ 

Pd--Cu  &  12$S^{2}_{0}\,(1-s)\,f^{Cu}_{16e}$ & 8.21 & 0.0030 & 
2.91 & 6.38 & 0.0025 & 2.90 & 5.57 & 0.0025 & 2.88   \\
(Pd--Pd$^{\prime}$) & 12$S^{2}_{0}\,(1-s)\,(1-f^{Cu}_{16e}$)  & & &\\ 

Pd--U   &   4$S^{2}_{0}\,(1-s)$  & 2.82 & 0.0025 & 
  3.05 & 2.32 & 0.0019 & 3.05 & 2.11 & 0.0027 & 3.05  \\

Pd$^{\prime}$--Cu  &  12$S^{2}_{0}\,s\,f^{Cu}_{16e}$ & 1.68 & 0.0025 & 4.31 & 
3.00 & 0.0036 & 4.31 & 3.41 & 0.0039 & 4.32   \\
(Pd$^{\prime}$--Pd$^{\prime}$) & 12$S^{2}_{0}\,s\,(1-f^{Cu}_{16e}$)  & & &\\ 

Pd$^{\prime}$--Cu$^{\prime}$  &  4$S^{2}_{0}\,s\,(1-f^{Pd}_{4c})$ & 
0.24 & 0.0051 & 4.58 & 0.35 &  0.0058 & 4.57 & 0.29 & 0.0065 & 4.55   \\
(Pd$^{\prime}$--U) & 4$S^{2}_{0}\,s$  & & &\\
(Pd$^{\prime}$--Pd) & 4$S^{2}_{0}\,s\,f^{Pd}_{4c}$  & & &\\

Pd--Cu  &  16$S^{2}_{0}\,(1-s)\,f^{Cu}_{16e}$ & 
10.95 & 0.0051 & 4.58 & 8.51 &  0.0058 & 4.57 & 7.43 & 0.0065 & 4.55   \\
(Pd--Pd$^{\prime}$) & 16$S^{2}_{0}\,(1-s)\,(1-f^{Cu}_{16e}$) & & &\\ 
\end{tabular*}
\end{ruledtabular}
\end{table*}

\begin{table}
\caption{Amount of Pd/Cu site interchange $s$ for various Pd
concentrations $x$ of \ucupdx{} determined from fits of the site 
interchange model to the Pd $K$
edge and U $L_{\textrm{III}}$ edge data.  These results from the U-edge data
are only reliable as a consistency check on the Pd-edge results (see 
Sec. \ref{Uedge}).  Errors in the last digit are given in parentheses.}

\label{svsxtable}
\begin{ruledtabular}
\begin{tabular}{ldd}
 $x$ & \multicolumn{2}{c}{$s$} \\\hline
    & \multicolumn{1}{c}{Pd $K$ edge} & \multicolumn{1}{c}{U
    $L_{\textrm{III}}$ edge} \\\hline
0.3 & 0.3(1) & 0.8(7) \\
0.5 & 0.25(3) & 0.4(4)\\
0.7 & 0.17(3) & 0.25(5)\\
0.9 & 0.26(4) & 0.35(9)\\
1.0 & 0.32(5) & 0.39(4)\\
1.25 & 0.38(4) & 0.40(8)\\
1.5 & 0.41(3) & 0.41(3)\\
\end{tabular}
\end{ruledtabular}
\end{table}

An alternative (equivalent) description of the site interchange from
$s$ and $x$ uses the occupancy fractions,  defined  as follows:
$f^{Pd}_{4c}$ and $f^{Cu}_{16e}$ are the fraction of $4c$ and $16e$
sites occupied by Pd and Cu, respectively, which are given by:
\begin{eqnarray*}
    f^{Pd}_{4c} = x(1-s)\\
    f^{Cu}_{16e}= 1-\frac{sx}{4}.
\end{eqnarray*}
These occupancy fractions are shown in Fig.  \ref{svsx}b.  The fits
are less sensitive to the parameter $f^{Cu}_{16e}$ due to the relative
abundance of $16e$ sites.  The amount of site interchange is roughly
constant up to $x$ = 0.7 at $s \sim 0.2$ and then increases
monotonically to a value of $s \sim 0.4$ at $x$ = 1.5.  Therefore, it
is possible that the system may be furthest away from random occupancy
of Pd at $x=0.7$, given that a minimum in $s$ exists at that
concentration.  For $x > 0.7$, an increasing number of Pd atoms are
located on the smaller $16e$ sites which is accompanied by an increase
of the lattice parameter.\cite{Chau_thesis,Andraka93} Note that the $
f^{Pd}_{4c}$ occupancy fractions exhibit a pronounced change in slope
close to $x=$0.7.

\subsection{U $L_{\textrm{III}}$ edge}
\label{Uedge}

The Pd/Cu site interchange determined from the Pd $K$ edge is
introduced into the model for the U $L_{\textrm{III}}$ edge data by
fixing the relative amplitudes of the various U--Cu and U--Pd bonds. 
Constraints on the fits to the U $L_{\rm III}$ data were chosen in a
similar manner to the Pd edge constraints, fitting in $r$-space to
pairs as long as 5.0 \AA, including 7 pairs but only 8 fit parameters. 
Representative fits to the data are shown in Fig.  \ref{ufits} and
summarized in Table \ref{ufitstable}.  The value of $S_{0}^{2}$ was
determined to be $0.80 \pm 0.05$.  In order to extract the static
continuous disorder, the square of the Debye Waller factor for the
U--Cu bond at 2.93 \AA{} $\sigma^2_{\rm U-Cu}(T)$ was fit to a
correlated-Debye model given by Eq. 
\ref{corr_debye}.\cite{Crozier88} The fits to this model are shown in
Fig.  \ref{utfits} for various concentrations of \ucupdx{} and the
results are collected in Table \ref{ufitstable}.  All measurements of
$\sigma^2_{\rm U-Cu}$ are consistent with zero disorder, as was
found in the Yb$X$Cu$_{4}$ ($X$ = Ag, Cd, In, Tl)
series.\cite{Lawrence01}

\begin{figure}
\includegraphics[width=3.3in]{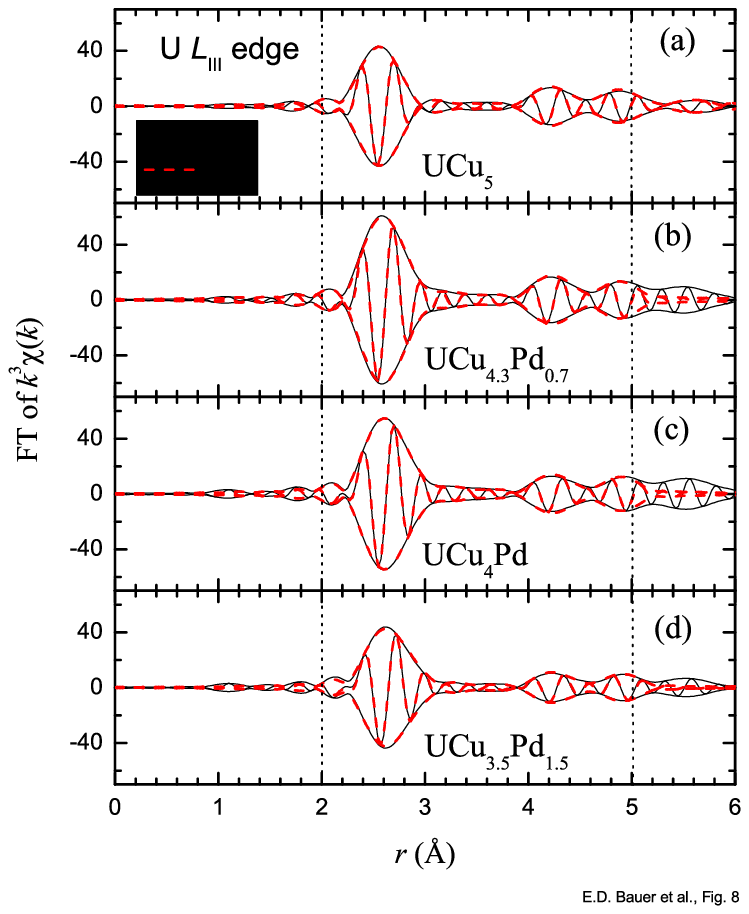}
\caption{Fourier transforms of $k^{3}\chi(k)$ of the U
$L_{\textrm{III}}$ edge data from selected \ucupdx{} samples.  
Solid lines are data for a) $x=0$ (15 K), b) $x=0.7$ (15 K), c) $x=1.0$ (3.3 K), and d) $x=1.5$ (15 K)
along with fits (dashed lines).
The fit range is from 2.0 to 5.0 \AA{} (dotted lines).}
\label{ufits}
\end{figure}

\begin{figure}
\includegraphics[width=3.3in]{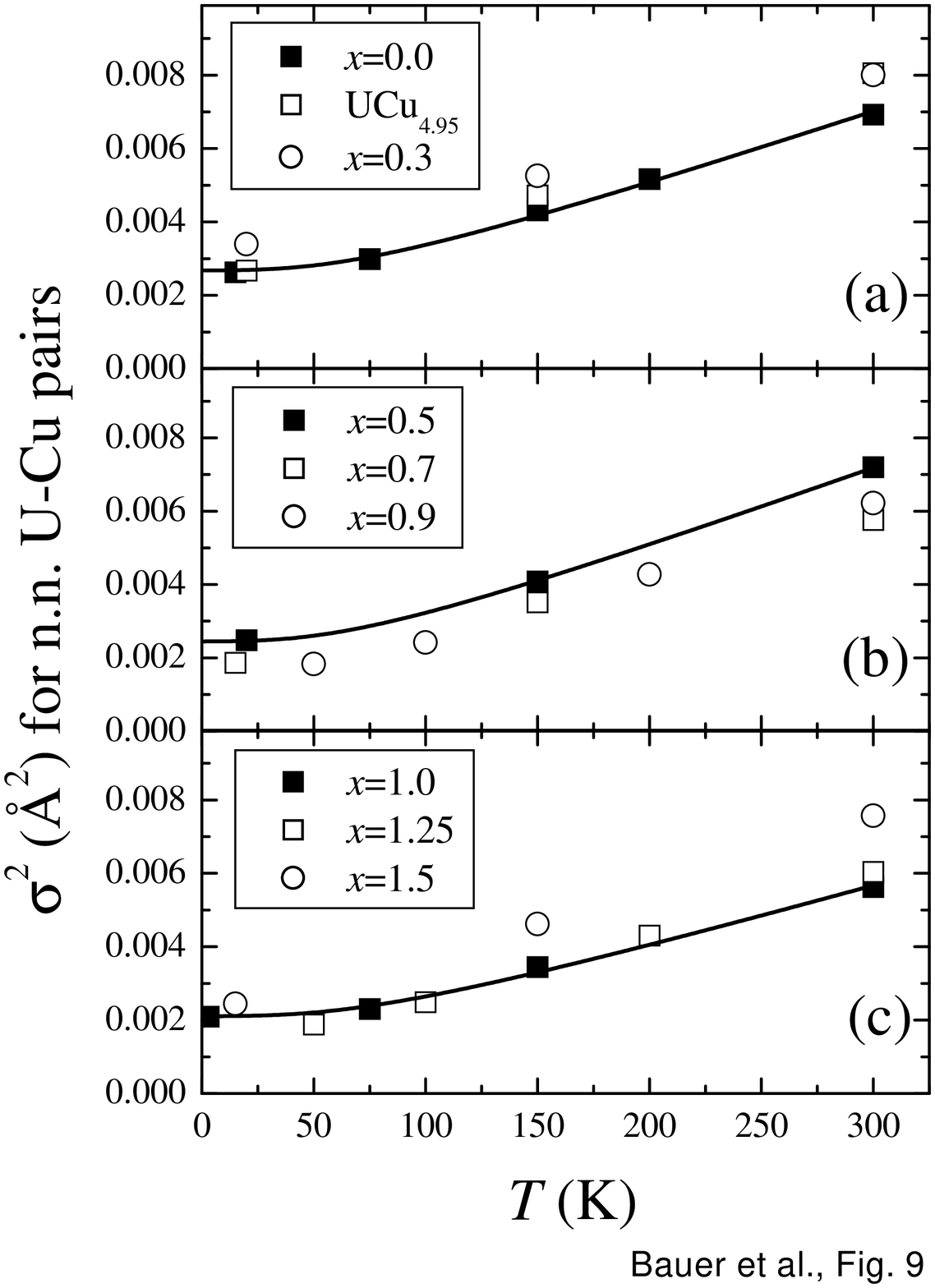}
\caption{Square of the Debye-Waller factor of the U-Cu bond at 2.93
\AA{} $\sigma^{2}_{\textnormal{U-Cu}}$ vs $T$ of \ucupdx{} determined
from the U $L_{\textrm{III}}$ edge fits.  Panel a): UCu$_{4.95}$,
$x=0$, and $x=0.3$; panel b) $x=0.5$, $x=0.7$, and $x=0.9$; panel b)
$x=1.0$, $x=1.25$, and $x=1.5$.  A typical fit to the correlated Debye
model is indicated by the solid line for $x=0$, $x=0.5$, and $x=1.0$
in panels a), b), and c), respectively.}
\label{utfits}
\end{figure}


\begin{table*}
\caption{Fit results of the site interchange model to some of the U
$L_{\textrm{III}}$ edge data of \ucupdx{}. Definitions and constraints are
similar to those in Table \ref{pdfitstable}. 
$S_{0}^{2}$
is determined to be 0.80 $\pm$0.05.}
\begin{ruledtabular}
\label{ufitstable}
\begin{tabular*}{\hsize}{lcddddddddd}
 \multicolumn{2}{l}{Pd Concentration $x$, $T$} &\multicolumn{3}{c}{0.7, 15 K}  & 
\multicolumn{3}{c}{1.0, 3.3 K} & \multicolumn{3}{c}{1.25, 15 K}\\
 Bond  &  $A$ constraint  & \multicolumn{1}{c}{$A$} &  
 \multicolumn{1}{c}{$\sigma^{2}$ (\AA$^{2}$)} & \multicolumn{1}{c}{$R$ (\AA)} &
\multicolumn{1}{c}{$A$} &  \multicolumn{1}{c}{$\sigma^{2}$ (\AA$^{2}$)} & 
\multicolumn{1}{c}{$R$ (\AA)} & \multicolumn{1}{c}{$A$} 
 & \multicolumn{1}{c}{$\sigma^{2}$ (\AA$^{2}$)} 
 & \multicolumn{1}{c}{$R$ (\AA)} \\\hline

U--Cu  & 12$S^{2}_{0}\,f^{Cu}_{16e}$ &9.31 &  0.0011 & 2.91 & 
8.83 & 0.0021 & 2.92 & 8.46 & 0.0019 & 2.93  \\

(U--Pd$^{\prime}$) &  12$S^{2}_{0}\,(1-f^{Cu}_{16e})$   \\

U--Cu$^{\prime}$ &  4$S^{2}_{0}\,f^{Pd}_{4c}$ & 1.86 & 0.0015 & 3.04 & 
2.18 & 0.0020 & 3.05 & 2.48 & 0.0042 & 3.07  \\

(U--Pd) &  4$S^{2}_{0}\,(1-f^{Pd}_{4c})$  \\

U--Cu &  16$S^{2}_{0}\,f^{Cu}_{16e}$ & 12.42 & 0.0027 & 4.56 & 
11.78 & 0.0035 & 4.57 & 11.28 & 0.0032 & 4.59  \\

(U--Pd$^{\prime}$) & 16$S^{2}_{0}\,(1-f^{Cu}_{16e})$   \\

U--U &   12$S^{2}_{0}$  & 9.60 & 0.0022 & 
  4.97 & 9.60 & 0.0028 & 4.99 & 9.60 & 0.0030 & 5.00  \\
\end{tabular*}
\end{ruledtabular}
\end{table*}

As a consistency check for the amount of site interchange, the U
$L_{\textrm{III}}$ edge data were fit with $s$ as a free parameter,
but subject to the relative amplitude constraints of the peaks given
by the site interchange model.  The values of $s$ obtained in this manner are 
only used to provide a qualitative comparison to those value determined 
from the Pd edge data since the  results of the U edge fits are much
more  model dependent (i.e. the U-Cu and U-Pd$^\prime$ pair distances are
constrained to each other)  and some 
parameters are strongly correlated with one another (i. e. $s$ and the near 
neighbor Debye-Waller factors). The results of these fits are
inconclusive for $x\leq 0.5$ due to the small amplitudes of the bonds
involving site interchange, but compare quite favorably with the
values obtained solely from the Pd edge for $x > 0.5$ (results listed
in Table \ref{svsxtable}).

\subsection{Cu $K$ edge}
\label{Cuedge}

The FT of $k^{3}\chi(k)$ of the Cu $K$ edge of \ucupdx{} for various
Pd concentrations is shown in Fig.  \ref{cu}.  The main peak at $\sim
2.3$ \AA{} is shifted to lower $r$ relative to the peak in the Pd $K$
edge data due to the larger number of Cu--Cu bonds at 2.50 \AA. The Cu
$K$ edge data were also fit by the site interchange model using
similar amplitude constraints involving $s$, $x$, and $S_{0}^{2}$. 
The value of $S_{0}^{2}$ is determined to be 0.61 $\pm$ 0.06.  Due to
the abundance of Cu present in the \ucupdx{} specimens, the fits are
not very sensitive to the site interchange parameter $s$ for the
relatively low Pd concentrations considered here.  The fits of this
model to the Cu $K$ edge data using the values of $s$ obtained from
the Pd $K$ edge are shown in Fig.  \ref{cu}.  
A quantitative comparison can be made to the Pd $K$
and U $L_{\textrm{III}}$ edge data for those peaks which dominate the
Cu $K$ edge spectra.  In general, there is good agreement between the
Cu edge fit results and the Pd and U edge results.  However, the details of the
fit parameters for some peaks with small amplitudes are only in
qualitative agreement with the fits to the other edge data.  For
instance, the Debye-Waller factors of the Cu--Pd$^{\prime}$ peak of
the $x=0.7$ sample ($\sigma^{2} = 0.0012$ \AA{}) and Pd$^{\prime}$--Cu
peak ($\sigma^{2} = 0.0029$ \AA{}) at 2.5 \AA{} as well as the bonds
lengths ($R= 2.48$ \AA{} for Cu--Pd$^{\prime}$ vs $R= 2.55$ \AA{} for
Pd$^{\prime}$--Cu) are somewhat different.  The difference in the
Debye-Waller factors is likely due to the particular choice of
constraints used in fitting the Cu edge data to the site interchange
model, i.e., the Debye-Waller factor of the Cu--Pd$^{\prime}$ peak was
set by the Cu--Cu peak which has the largest amplitude
in the XAFS spectra (see Sec. \ref{Pdedge}).  Therefore, the Debye-Waller factor of the Cu-Cu
peak dominates the fit and its small magnitude could be due to the
presence of small amounts of free Cu in some samples causing a lower
(incorrect) value for $S_{0}^{2}$ for the Cu edge data.  A plausible
explanation for the difference in bond lengths of the nearest neighbor
bonds is a slight distortion of the Cu tetrahedra as 
discussed below.  Otherwise, the fit parameters obtained from the Cu
$K$ edge data are consistent with the other edges and so are not
reported in detail here.

\begin{figure}
\includegraphics[width=3.3in]{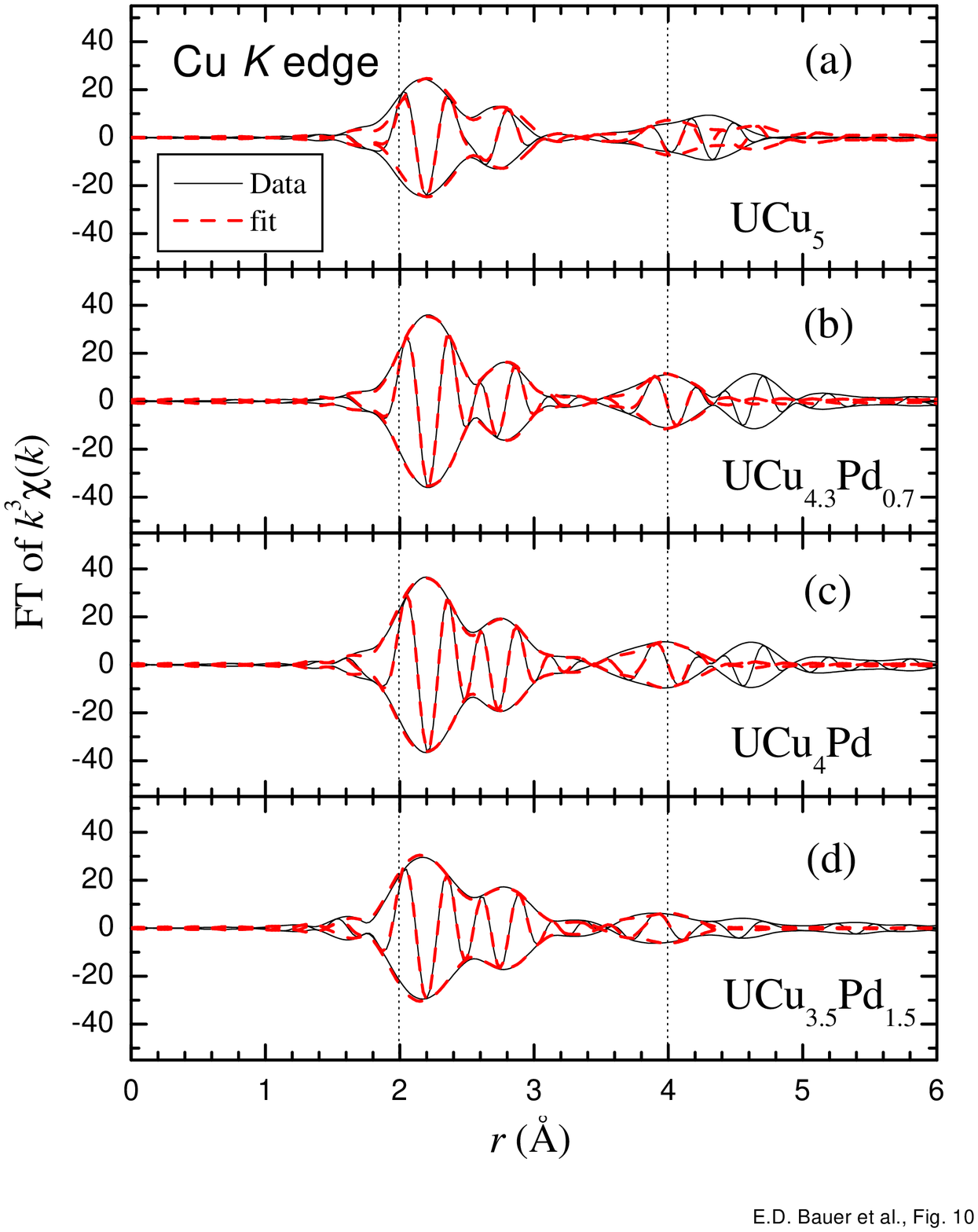}
\caption{Fourier transforms of $k^{3}\chi(k)$ of the Cu $K$ edge data from 
selected \ucupdx{} samples.  Solid lines are data for a)
$x=0$ (15 K), b) $x=0.7$ (20 K), c) $x=1.0$ (20 K), and d) $x=1.5$ (15 K) along 
with fits (dashed lines).  The fit range is from 2.0 to 4.0 \AA{} (dotted lines).}
\label{cu}
\end{figure}

\section{Consequences of Measured Lattice Disorder on $V$ and $T_{K}$}
\label{effects}

The distribution of hybridization strengths $P(V)$ for \xlow{} is
shown in Fig.  \ref{vdistribution} using the measured value of
$\sigma_{static}^{2}=0.0002$ \AA$^2$ obtained from the U
$L_{\textrm{III}}$ edge data along with the curves for one standard
deviation of this value ($\sigma_{static}^{2}=0$ \AA$^2$ (binomial)
and $\sigma_{static}^{2}=0.00078$ \AA$^2$).  The value of
 $\sigma_{static}^2$=$\sigma_{KDM}^2$  (=0.00343 \AA$^2$) necessary to fit the $\chi(T)$
data with the KDM model produces a distribution that is considerably
broader than the one derived from the measured amount of static
disorder.  The mean hybridization $<V>$ for all Pd concentrations was
calculated using the measured values of $s$ and $\sigma_{static}^2$ and
is shown in Fig.  \ref{vdistribution}a along with the spread in
hybridization strengths $\Delta V/<V>$ (Fig.  \ref{vdistribution}b). 
The average hybridization strength increases linearly with increasing
Pd concentration reflecting the additional hybridization due to the
larger Pd atoms, while the width of the distribution increases only
slightly.

\begin{table}
\begin{ruledtabular}
\label{sigmas}
\caption{Comparison of measured and calculated static continuous
disorder in \ucupdx.  The parameters $\sigma_{static}^{2}$ and
$\theta_{cD}$ were determined from fits  to  the U
$L_{\textrm{III}}$ edge Debye-Waller factors to a correlated Debye
model.  The values of the static disorder
necessary to fit the magnetic susceptibility $\sigma_{KDM}^2$ are
determined from fits of the Kondo disorder model to the $\chi(T)$ data.
Errors in the last digit are given in parentheses.}
\begin{tabular*}{\hsize}{l|dldd}
$x$ & \multicolumn{1}{c}{$\sigma^{2}_{U-Cu}$}  &
 \multicolumn{1}{c}{$\theta_{cD}$} & 
 \multicolumn{1}{c}{$\sigma^{2}_{static}$ } & 
 \multicolumn{1}{c}{$\sigma_{KDM}^{2}$} \\
 & \multicolumn{1}{c}{(\AA$^{2}$)}& (K) & \multicolumn{1}{c}{(\AA$^{2}$)}
 & \multicolumn{1}{c}{(\AA$^{2}$)} \\\colrule
UCu$_{4.95}$  & 0.0027(3) & 290(4) & 0.0004(3) &    \\

0 & 0.0026(3) & 318(7) & 0.0006(3) &    \\

0.3 & 0.0034(3) & 310(8) & 0.0013(3) &    \\

0.5 & 0.0025(4) &  306(10) &  0.0003(4) &    \\

0.7 & 0.0011(3) &  335(15) &  0.0000(3) &    \\

0.9 & 0.0018(4) &  314(2) &  -0.0005(4) &  0.0038  \\

1.0 & 0.0021(6) & 345(7) &  0.0002(6) &  0.0030  \\

1.25 & 0.0019(3) &  322(4) &  -0.0003(3) &  0.0028  \\

1.5 &  0.0023(6) &  300(7) &  0.0003(6) &  0.0028  \\
\end{tabular*}
\end{ruledtabular}
\end{table}

\begin{figure}
\includegraphics[width=3.3in]{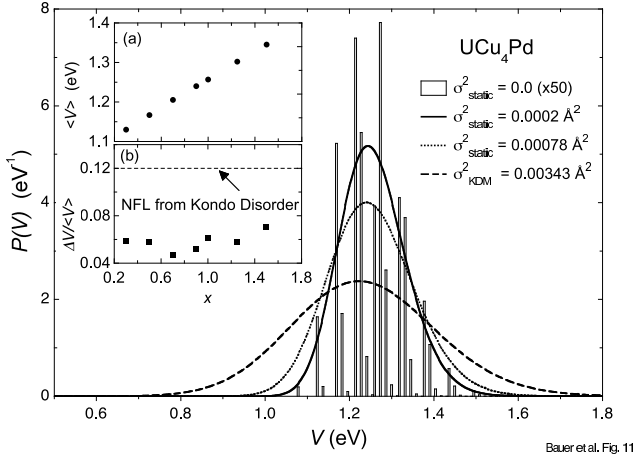}
\caption{Distribution of hybridization strengths $P(V)$ for
UCu$_{4}$Pd.  The solid curve is the distribution using the measured
value of $\sigma_{static}^2$ obtained from the U $L_{\textrm{III}}$ edge
data along with the distributions within one
standard deviation of the measured value of $\sigma_{static}^2$ (dotted
line).  The dashed line uses the value of $\sigma_{KDM}^2$ necessary to
fit the $\chi(T)$ data with the KDM model.  Inset a): Average
hybridization $<V>$ vs $x$.  Inset b): Spread in hybridization
strengths $\Delta V/<V>$ vs $x$.}
\label{vdistribution}
\end{figure}

The values of the square of the Debye-Waller factor
$\sigma_{static}^{2}$, determined from the U $L_{\textrm{III}}$ edge
data for the U-Cu bond at 2.93 \AA{}, are compared to those values of
the static disorder needed to fit the $\chi(T)$ data using the KDM and
are displayed in Fig.  \ref{sigmastatic}.  The amount of the measured
continuous static disorder is small and consistent with zero disorder. 
The value of $\sigma_{static}^{2}$ for $x=0.3$ is larger than for the
other concentrations.  This anomaly may be attributed to the influence
of sample oxidation since no special care was taken to minimize
oxidation in this compound as was the case for UCu$_{5}$ and
UCu$_{4.95}$ samples.  However, no trace of UO$_2$ was found in the U
$L_{\textrm{III}}$ XAFS spectra. The amount of static disorder used to
reproduce the logarithmic divergence in the magnetic susceptibility
with the KDM $\sigma_{KDM}^{2}$ is indicated by the shaded region in
Fig.  \ref{sigmastatic}, reflecting the uncertainty in comparing the
model with the experimental data.  The lower bound on
$\sigma_{KDM}^{2}$ is estimated to be 20\%  below  the smallest value of
$\sigma_{KDM}^{2}$ ($x=1.5$).

\begin{figure}
\includegraphics[width=3.3in]{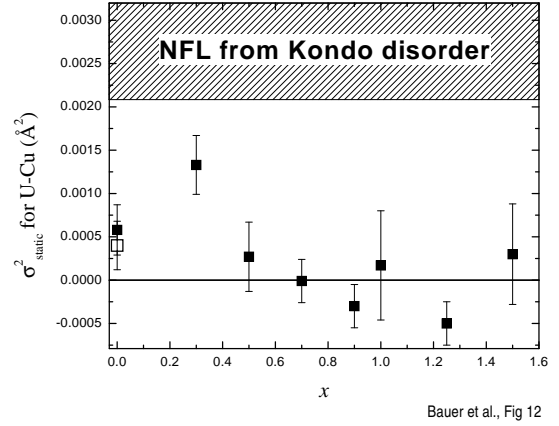}
\caption{The static part of the bond-length distribution variance for the 
U-Cu bond at $\sim$2.93 \AA{}, $\sigma^{2}_{\textrm{static}}$ vs $x$ of 
\ucupdx. Data for the UCu$_{4.95}$ sample is denoted by the open square.  
The shaded area is the approximate amount of disorder necessary
 to fully explain NFL behavior in \ucupdx{}
compounds based on a distribution of hybridization strengths generated by lattice disorder .}
\label{sigmastatic}
\end{figure}

\section{Discussion}
\label{discussion}

In  our  previous XAFS study on UCu$_{4}$Pd, Pd/Cu site interchange was found to occur, 
providing evidence that {\textit{chemical}}  disorder  exists in this compound.\cite{Booth98c} 
 We 
suggested from these results that lattice disorder  is the microscopic origin 
of the Kondo disorder model and therefore, produces  the observed NFL characteristics. 
However, a necessary assumption for this lattice origin of the KDM was the inclusion of 
   bond-length disorder (at that time unquantifiable), presumably caused by the 
site interchange.  This assumption is 
tested in the present investigation by employing  an experimental 
determination of the static continuous disorder about the U atoms (Fig. \ref{sigmastatic}) 
for a number of different Pd concentrations.  These $\sigma_{static}^{2}$ results are obtained from a 
 more sophisticated fitting  model of the local Pd environment and are compared to those values 
of $\sigma_{KDM}^{2}$ necessary to explain the NFL properties of \ucupdx{}  ($x=0.9-1.5$)
via the Kondo disorder model.   This comparison was not possible in the original 
work\cite{Booth98c} because the fitting model was not complete enough and there were not any 
data on other Pd concentrations, making a determination of certain relevant 
experimental parameters (especially $S_0^2$) less reliable. 
Our results in the present study indicate that while Pd/Cu site interchange exists for all
Pd concentrations measured, it does not produce enough static
continuous disorder to explain the non-Fermi liquid behavior observed
in the magnetic susceptibility (and presumably the specific heat)
{\textit{within the variant of the Kondo disorder model discussed here}, i.e. the KDM with 
a distribution of $T_{K}$s produced by lattice disorder in the hybridization 
strength $V$. 
It is important to emphasize this last caveat since other sources of disorder aside from a 
distribution of hybridization strengths due to bond length disorder, such
as a distribution of the density of states or $f$-ion energies,  could cause a 
distribution of Kondo temperatures. Although the relationship between the
\textit{local} density of states (the relevant quantity) and that
derived from band structure calculations is unclear since these
calculations are usually done in the clean limit (i.e., no disorder),
fluctuations in the local density of states are a real concern given
the shape of the $d$-level density with energy expected from band
structure calculations on YbCu$_4$Pd.\cite{Monachesi96}   Moreover, the tight-binding
model for the hybridization treats all the near neighbor U-Cu atom pairs equivalently, and therefore
does not include potential phase differences that may occur in the presence of
disorder, possibly broadening the distribution of $T_{K}$s.\cite{Cox02} 
Therefore, the Kondo disorder model may still apply to \ucupdx{} if other 
disorder effects are the dominant sources of  Kondo disorder other than a 
distribution of hybridization strengths due to bond length disorder.
An
explanation for the presence of site interchange, yet lack of a spread
in U--M bond lengths, is that site interchange causes a distortion in
the Cu tetrahedra which is roughly perpendicular to the
U--Pd$^{\prime}$ or U--Cu bonds.  Support for this scenario is
provided by the difference in the bond lengths of the
Pd$^{\prime}$--Cu ($R \sim 2.54$ \AA) and Cu--Pd$^{\prime}$ ($R \sim
2.48$ \AA) observed for all Pd concentrations.  The variation in the
U--Pd$^{\prime}$ bond length due to this distortion is small, i.e.,
$\Delta R \sim [(2.93 \, {\textrm \AA})^{2} + (0.06 \,{\textrm
\AA})^{2}]^{1/2} - 2.93 \,{\textrm \AA} \sim 0.0006\, {\textrm \AA}$.

  A recent
calculation that incorporates local fluctuations in the density of
states, $\epsilon_{f}$, and $V$ into a disordered Anderson lattice
model with $\approx$ 200 sites has been performed by Miranda and
Dobrosavljevi{\'{c}}.\cite{Miranda01a,Miranda01b} Their results
indicate the presence of Griffiths' singularities (not associated with
the proximity to magnetic order\cite{CastroNeto98,CastroNeto00}),
giving rise to NFL behavior near a metal-insulator transition; the NFL
behavior was found to exist for all three types of disorder mentioned
above over a broad range of values of those
parameters.\cite{Miranda01a,Miranda01b} In this model, the presence of
disorder will cause a variation in the magnitude of the conduction
electron wavefunctions leading to a spread in the density of states. 
Therefore, even a discrete distribution of hybridization strengths due
to site interchange alone (or a small amount of continuous disorder)
will, in principle, give rise to Griffiths' singularities and NFL
behavior within this disordered Anderson lattice model.  However,
calculations within this model starting from a discrete distribution
of hybridizations similar to that in Fig.  \ref{vdistribution} do not
generate enough width to the density of states to explain the NFL
behavior.\cite{Miranda_priv}

While there is some supporting evidence for the applicability of the
Kondo disorder model to \ucupdx,\cite{Bernal95,MacLaughlin98} more
recent studies,\cite{MacLaughlin01,Aronson01,Weber01,Buttgen00} in
addition to this one, suggest other non-Fermi liquid models may be
more appropriate.  For instance, even though the specific
heat of annealed samples of UCu$_{4}$Pd exhibit a NFL-like logarithmic 
divergence, the electrical resistivity no longer displays a linear 
$T$-dependence, clearly at
odds with the predictions of the Kondo disorder model.\cite{Weber01} 
In addition, while the Cu spin-lattice relaxation rates $1/T_{1}$ of 
UCu$_{3.5}$Pd$_{1.5}$ in magnetic 
fields below $H=5$ T are consistent with the KDM,\cite{Ambrosini99,Buttgen00}
$1/T_{1}$ in higher magnetic fields up to $H=9$ T 
is not well described by this model.\cite{Buttgen00}
   Recent $\mu$SR
measurements\cite{MacLaughlin01} on UCu$_{4}$Pd and
UCu$_{3.5}$Pd$_{1.5}$ reveal that the muon relaxation rates are two
orders of magnitude faster than what is expected from the KDM. There
is growing evidence from NMR,\cite{MacLaughlin98},
$\mu$SR,\cite{MacLaughlin01} and inelastic neutron scattering
experiments\cite{Aronson01} that short-range magnetic correlations
(less than a unit cell distance) are present in \ucupdx.  One possible
explanation for these magnetic correlations is the formation of
magnetic clusters in the vicinity of a quantum critical point in a
disordered  material, similar to the scenario  proposed by Castro Neto and 
coworkers.\cite{CastroNeto98,CastroNeto00} The spin glass behavior
observed\cite{Scheidt98,Vollmer00} in the $x=1$ and $x=1.5$ samples at
$T_{f} \sim 0.2-0.3$ K could be due to the freezing of these magnetic
clusters in this picture.  The power law behavior of the specific heat
and magnetic susceptibility of  such  compounds and other Ce- and
U-based materials\cite{deAndrade98,Vollmer00} is consistent with this
Griffiths' phase model.  However, the Griffiths' phase predictions of
the dynamics of the magnetic correlations do not agree with the
inelastic neutron and $\mu$SR measurements, which are perhaps best
viewed within a framework of a quantum critical point scenario in
which spin glass\cite{Sachdev98,Grempel99} or antiferromagnetic
order\cite{Hertz76,Millis93,Millis95,Continentino94} has been suppressed to $T=0$
K. Moreover, it is not clear whether the observed magnetic
correlations\cite{Aronson01} are strong enough to allow for the
Griffiths' phase model.  Clearly, the theoretical understanding of
this system is far from complete.

An investigation\cite{Weber01} of the effects of annealing on the
physical properties of UCu$_{4}$Pd was recently reported by Weber and
coworkers and provides another way to study the role of disorder in
this system.  Annealing the samples at 750 $^{\circ}$C for 1 and 2
weeks produced a decrease in the lattice parameter $a$, more
consistent with the nearly linear
relation\cite{Andraka93,Chau_thesis,Korner00,Weber01} observed (for unannealed
samples) between $a$ and $x$ for $x < 0.7$, while splat quenching the
sample results in an increase of $a$, relative to the unannealed
sample.  One possible interpretation of this behavior is that the
unannealed sample has a comparable amount of site interchange to the
one studied here and splat quenching gives rise to more site
interchange while annealing reduces it.  However, note that the 
lattice parameter of the annealed samples extrapolates to the linear
$a$ vs $x$ relation of the unannealed samples below $x$=0.7.  Since
the unannealed samples have site interchange, even for low Pd concentrations,
the annealed $x$=1.0 specimens most likely also have a
considerable amount of site interchange.  In any case, even though the
specific heat divided by temperature of the annealed samples shows a
logarithmic dependence over a larger temperature range (0.08-10 K)
compared to the unannealed specimens, the electrical resistivity no
longer exhibits a linear $T$-dependence, indicating that the low
temperature properties are greatly influenced by disorder.  Our
results on unannealed samples as shown in Fig. \ref{svsx}
indicate an increase in the site
interchange for samples above $x = 0.7$.  Therefore, these XAFS
results are consistent with the interpretation of Weber {\textit{et
al.}}\cite{Weber01} that annealing reduces the site interchange, although
it does not suppress it completely.  Future work, such as XAFS, $\mu$SR,
and neutron scattering measurements on annealed samples of \ucupdx{}
should continue to provide information that will lead to a better
understanding of the interplay between disorder and the NFL properties
in this system.

\section{Conclusions}
\label{conclusions}
We have measured the local structure about the U, Cu,
and Pd atoms in \ucupdx{} (0$ \leq x \leq 1.5$) using the XAFS
technique.  A model involving Pd/Cu site interchange was used to fit
the Pd and Cu $K$ edge and U $L_{\textrm{III}}$ edge data.  Pd/Cu site
interchange $s$ was found to occur in all samples with a roughly
constant value of $s \sim 0.2$ up to $x =0.7$ and increased to
$s \sim 0.4$ at $x=1.5$.  These results were used to determine the
static disorder about the U atoms.  The magnetic susceptibility of
\ucupdx{} was calculated with the Kondo disorder model via a tight
binding approximation for the hybridization strength $V_{fd}$,
assuming that lattice disorder is the cause of Kondo disorder.  Our
results indicate that the measured static disorder due to Pd/Cu site
interchange  does not produce a sufficient width to the distribution of (tight-binding) $V_{fd}$
to generate NFL behavior within the Kondo disorder model 
 and suggests either  that there are sources of Kondo disorder other than 
a distribution of hybridization strengths with a 
lattice-disorder origin or that the Kondo disorder model 
is not applicable to \ucupdx.  
 However, the presence
of significant Pd/Cu site interchange should in principle still cause
a distribution of hybridization strengths and may well couple to
changes in the local electron charge densities, and therefore strongly
favor NFL models which include disorder, such as the various Griffiths'
phase models.

\begin{acknowledgments}
We thank Prof.  F. Bridges for assistance in collecting the XAFS data.  We
also to thank D. E. MacLaughlin, J. M. Lawrence, D. L. Cox, E. Miranda,
and E.-W. Scheidt for many useful discussions.  This work was partially
supported by U. S. National Science Foundation under Grant No. 
DMR00-72125, the U. S. Department of Energy (DOE) under Grant No.  DE
FG03-86ER-45230, and by the Office of Basic Energy Sciences (OBES),
Chemical Sciences Division of the DOE, Contract No.  AC03-76SF00098. 
XAFS data were collected at the Stanford Synchrotron Radiation
Laboratory, which is operated by the DOE/OBES.
\end{acknowledgments}

\bibliographystyle{apsrev}
\bibliography{bibli}

\end{document}